\documentclass[pra,onecolumn,preprint,titlepage]{revtex4}
\usepackage{graphicx}
\usepackage{amsfonts}

\begin{document}

\preprint{}
\title[Optical analogies to quantum states]{Optical analogies to quantum
states and coherent detection of pseudorandom phase sequence}
\author{Jian Fu, Wenjiang Li and Yongzheng Ye}
\affiliation{State Key Lab of Modern Optical Instrumentation, College of Optical Science
and Engineering, Zhejiang University, Hangzhou, 310027, China}

\begin{abstract}
The key to optical analogy to a multi-particle quantum system is the
scalable property. Optical fields modulated with pseudorandom phase
sequences is an interesting solution. By utilizing the properties of
pseudorandom sequences, mixing multiple optical fields are distinguished by
using coherent detection and correlation analysis that are mature methods in
optical communication. In this paper, we utilize the methods to investigate
optical analogies to multi-particle quantum states. In order to demonstrate
the feasibility, numerical simulations are carried out in the paper, which
is helpful to the experimental verification in the future.
\end{abstract}

\date{today}
\startpage{1}
\email{jianfu@zju.edu.cn}
\maketitle

\section{Introduction \label{sec1}}

Quantum computation is a revolutionary approach to speed up many classical
algorithms exponentially, due to quantum superposition and entanglement \cite%
{Nielsen,Jozsa}. However, it is difficult to construct a powerful enough
quantum computer to implement sufficiently complex quantum algorithms, which
inspires some researches on the simulations of quantum states and quantum
computation using classical optical fields \cite%
{Spreeuw,Toppel,Qian,Qian2,Benjamin}. In these researches, the simulations
can be realized using classical optical fields to introduce extra freedoms,
such as orbital angular momentum, frequency, and time bins \cite{Sandeep},
which, however, might be hard to simulate a quantum state with arbitrary
number of quantum particles.

Recently, a method is proposed to simulate quantum states and quantum
computation using the classical optical fields modulated by pseudorandom
phase sequences \cite{Fu1,Fu2,Fu3,Fu4}. In the method, two orthogonal modes
of the optical fields, such as polarization and spatial mode \cite{Fu5}, are
encoded as qubit $\left\vert 0\right\rangle $ and $\left\vert 1\right\rangle 
$, which demostrate the similar properties to quantum states, such as
superposition even entanglement. However, the physical meaning of the
superpositions is different from quantum states. A measurement of certain
quantum states will generally produce a random result, and the ensemble
averaging after many measurements can yield a physical measurement result.
In Ref. \cite{Fu3}, a new conception of pseudorandom phase ensemble is
introduced. In the theoretical framework, the nonlocal properties of quantum
entanglement can be simulated by using classical fields modulated with
pseudorandom phase sequences \cite{Fu1}, which contain all properties of the
coherent superposition of orthogonal modes and the orthogonality, closure
and balance of pseudorandom phase sequences \cite{Viterbi,Peterson,PSK}. It
is interesing to make analogies between a single quantum particle with a
phase code in a pseudorandom phase sequence, and quantum ensemble with
pseudorandom phase sequences. From these analogies, we could conclude that
pseudorandom phase sequences not only can qualify independence and
distinguishability for each classical field, also can provide similar
randomness to quantum measurements. Moreover, the randomness is ergodic
within one sequence cycle, which means a higher efficiency than the
ergodicity of real quantum measurement system. Furthermore, the states with
arbitrary number of quantum particles can be simulated by classical fields
with same number of pseudorandom sequences whose number increase linearly
with their length. It means the simulation is effective and without
limitation.

In this paper, we discuss the coherent detection and correlation analysis
method to distinguish the optical fields modulated with pseudorandom phase
sequences, and then to simulate the multi-particle quantum states, such as
GHZ state and W state. Furthermore, we construct an optical analogy to the
entangled state as the result of the modular exponential function in Shor's
algorithm \cite{Shor} to factorize $15=3\times 5$. We demonstrate the result
state how to be represented by classical fields and measured by the coherent
detection and correlation analysis. The computer simulation software adopted
in this paper is the well-known optical communication simulation software
OPTISYSTEM.

\section{Orthogonality of pseudorandom phase sequence and coherent detection
of optical fields \label{sec2}}

In modern communication, the pseudorandom code is widely used in CDMA (Code
Division Multiple Access) to distinguish different users \cite%
{Viterbi,Peterson,PSK}. The orthogonality of the code can enable the
coherent detection in communication, which can transfer target information
to the users with a corresponding code in a same channel used by many users.
Ref. \cite{Fu2} proposed that quantum entanglement and quantum state can be
simulated by classical optical fields modulated by this pseudorandom phase
sequences, which make use of the coherent superposition of classical optical
fields and the orthogonality, closure and balance property of pseudorandom
phase sequences. Especially, the simulation of the non-locality in quantum
mechanism is essentially utilize the non-locality of the phase of classical
fields \cite{Fu1,Fu3}. These properties can be demonstrated by the coherent
detection of classical optical fields. Here, we will utilize the mature
optical coherent communication technology and the simulation software,
OPTISYSTEM, to investigate the orthogonality and indistinguishability of
multi-optical-field realized by orthogonal pseudorandom phase sequences.

\begin{figure}[tbph]
\centering\includegraphics[width=5.047in]{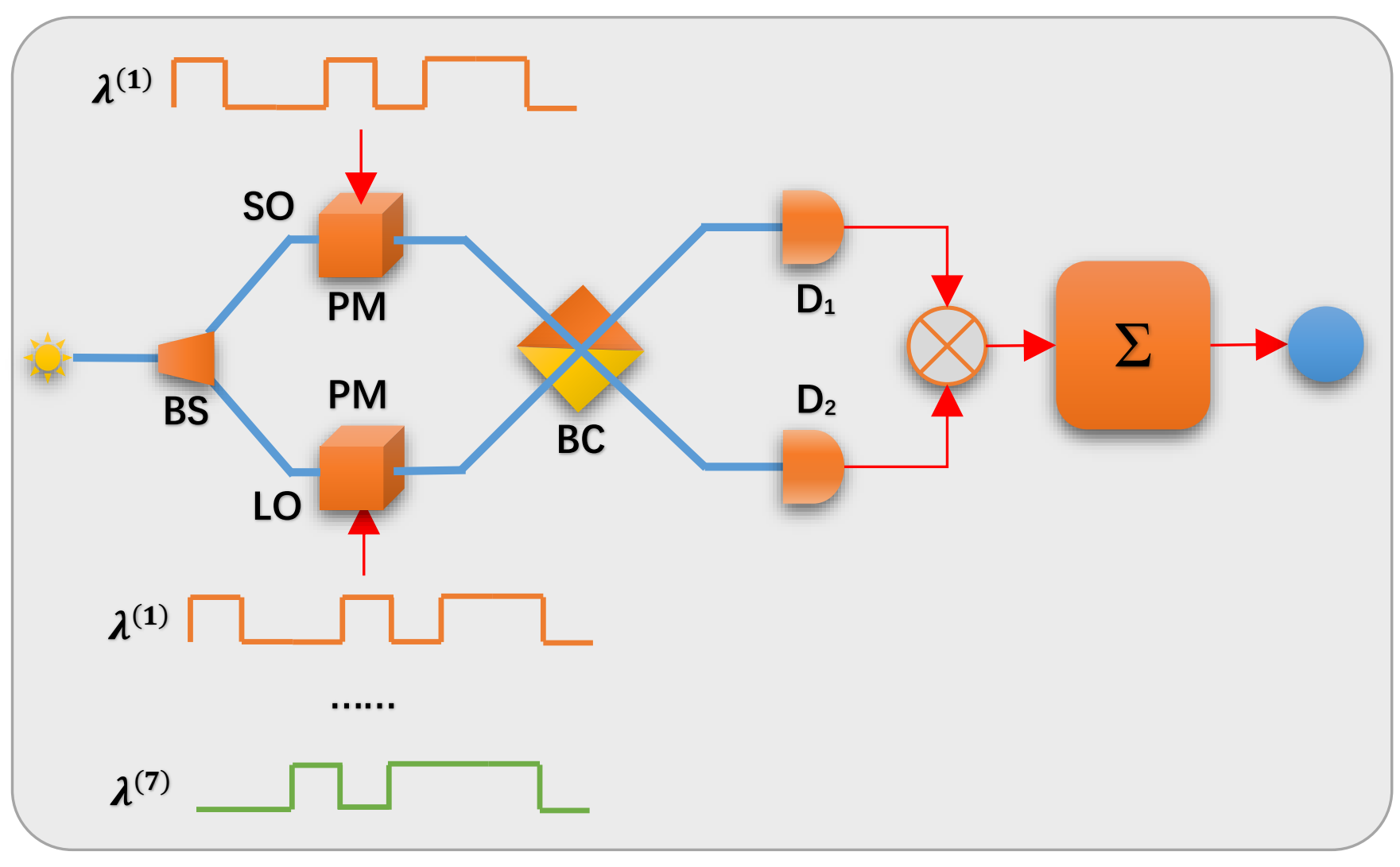}
\caption{The scheme of the coherent detection of pseudorandom phase
sequence, where SO: the signal light, LO: the local light, BS: beam
splitter, PM: phase modulator, BC: beam coupler, $D_{1}$ and $D_{2}$:
photodetectors, $\otimes $: multiplier and $\Sigma $: integrator (integrate
over entire sequence period).}
\label{1}
\end{figure}

According to Ref. \cite{Fu2}, we encode two orthogonal polarization modes of
classical fields as quantum bits (qubits) $\left\vert 0\right\rangle $\ and\ 
$\left\vert 1\right\rangle $. In order to distinguish different classical
fields, we modulated the fields with pseudorandom phase sequences. These
pseudorandom phase sequences, except the all-zero $\lambda ^{\left( 0\right)
}$ sequence, are the expressions as following,%
\begin{eqnarray}
\lambda ^{\left( 1\right) } &=&\left\{ 
\begin{array}{cccccccc}
\frac{\pi }{2} & 0 & 0 & \frac{\pi }{2} & 0 & \frac{\pi }{2} & \frac{\pi }{2}
& 0%
\end{array}%
\right\} ,  \label{e1} \\
\lambda ^{\left( 2\right) } &=&\left\{ 
\begin{array}{cccccccc}
\frac{\pi }{2} & \frac{\pi }{2} & 0 & 0 & \frac{\pi }{2} & 0 & \frac{\pi }{2}
& 0%
\end{array}%
\right\} ,  \nonumber \\
\lambda ^{\left( 3\right) } &=&\left\{ 
\begin{array}{cccccccc}
\frac{\pi }{2} & \frac{\pi }{2} & \frac{\pi }{2} & 0 & 0 & \frac{\pi }{2} & 0
& 0%
\end{array}%
\right\} ,  \nonumber \\
\lambda ^{\left( 4\right) } &=&\left\{ 
\begin{array}{cccccccc}
0 & \frac{\pi }{2} & \frac{\pi }{2} & \frac{\pi }{2} & 0 & 0 & \frac{\pi }{2}
& 0%
\end{array}%
\right\} ,  \nonumber \\
\lambda ^{\left( 5\right) } &=&\left\{ 
\begin{array}{cccccccc}
\frac{\pi }{2} & 0 & \frac{\pi }{2} & \frac{\pi }{2} & \frac{\pi }{2} & 0 & 0
& 0%
\end{array}%
\right\} ,  \nonumber \\
\lambda ^{\left( 6\right) } &=&\left\{ 
\begin{array}{cccccccc}
0 & \frac{\pi }{2} & 0 & \frac{\pi }{2} & \frac{\pi }{2} & \frac{\pi }{2} & 0
& 0%
\end{array}%
\right\} ,  \nonumber \\
\lambda ^{\left( 7\right) } &=&\left\{ 
\begin{array}{cccccccc}
0 & 0 & \frac{\pi }{2} & 0 & \frac{\pi }{2} & \frac{\pi }{2} & \frac{\pi }{2}
& 0%
\end{array}%
\right\} ,  \nonumber
\end{eqnarray}%
where the sequences are GF(2) with $\left[ 0,\frac{\pi }{2}\right] $ instead
of GF(4) with $\left[ 0,\frac{\pi }{2},\pi ,\frac{3\pi }{2}\right] $ in Ref. 
\cite{Fu1}. Due to the orthogonality of polarization modes, the pseudorandom
phase sequences only need to distinguish the classical fields with same
polarization mode. In this section, we mainly focus on the characteristic of
the classical fields with same polarization mode after the modulation of
pseudorandom phase sequences. We choose $\lambda ^{\left( 1\right) }$ to
modulate the classical fields labeled as signal light (SO), the electric
field component of the field is:%
\begin{equation}
E_{S}\left( t\right) =A_{S}e^{-i\left( \omega t+\lambda _{k}^{\left(
1\right) }\right) },  \label{e2}
\end{equation}%
where $A_{S},\omega $ are the amplitude and frequency of the classical
optical field respectively, and $\lambda _{k}^{\left( 1\right) }$ is the
phase code of $\lambda ^{\left( 1\right) }$ at time $t$. In order to do the
coherent detection of pseudorandom phase sequence, we design a detection
scheme shown in Fig. \ref{1}, according to the method used in coherent
optical communication \cite{Viterbi,Peterson,PSK}. The detection scheme
makes the local light (LO) and signal light (SO) interfere with each other.
In order to ensure the coherence of them, these two beams are obtained by
splitting the same source by a beam splitter. The field of local light can
be expressed as: 
\begin{equation}
E_{L}\left( t\right) =A_{L}e^{-i\left( \omega t+\lambda _{k}^{\left(
n\right) }\right) },  \label{e3}
\end{equation}%
where $\lambda ^{\left( n\right) }$ can be arbitrary sequences in Eq. (\ref%
{e1}) and the amplitude $A_{L}=A_{S}$. After the coherent superposition
throught the coupler, the ouput lights can be expressed respectively, 
\begin{equation}
\left( 
\begin{array}{c}
E_{1}\left( t\right) \\ 
E_{2}\left( t\right)%
\end{array}%
\right) =\frac{1}{\sqrt{2}}\left( 
\begin{array}{cc}
1 & i \\ 
-i & 1%
\end{array}%
\right) \left( 
\begin{array}{c}
E_{S}\left( t\right) \\ 
E_{L}\left( t\right)%
\end{array}%
\right) =\frac{A_{S}}{\sqrt{2}}\left( 
\begin{array}{c}
e^{-i\left( \omega t+\lambda _{k}^{\left( 1\right) }\right) }+ie^{-i\left(
\omega t+\lambda _{k}^{\left( n\right) }\right) } \\ 
-ie^{-i\left( \omega t+\lambda _{k}^{\left( 1\right) }\right) }+e^{-i\left(
\omega t+\lambda _{k}^{\left( n\right) }\right) }%
\end{array}%
\right) .  \label{e4}
\end{equation}%
Then, the output electic signals of photodetectors (PDs) $D_{1}$ and $D_{2}$
is proportional to 
\begin{eqnarray}
I_{1} &=&\mu \left\vert E_{1}\left( t\right) \right\vert ^{2}=\mu A_{s}^{2} 
\left[ 1+\sin \left( \lambda _{k}^{\left( 1\right) }-\lambda _{k}^{\left(
n\right) }\right) \right] ,  \label{e5} \\
I_{2} &=&\mu \left\vert E_{2}\left( t\right) \right\vert ^{2}=\mu A_{s}^{2} 
\left[ 1-\sin \left( \lambda _{k}^{\left( 1\right) }-\lambda _{k}^{\left(
n\right) }\right) \right] ,  \nonumber
\end{eqnarray}%
where $\mu $ is the parameter related to the sensitivity of PDs. Finally,
after the correlation analysis of two electric signal, we can obtain the
following according to the orthogonality of pseudorandom sequence, 
\begin{equation}
C=\left\langle I_{1}I_{2}\right\rangle =\frac{\mu ^{2}A_{s}^{4}\Delta T}{2}%
\sum\limits_{k=1}^{8}\left[ 1+\cos 2\left( \lambda _{k}^{\left( 1\right)
}-\lambda _{k}^{\left( n\right) }\right) \right] =\left\{ 
\begin{array}{c}
8\mu ^{2}A_{s}^{4}\Delta T,n=1 \\ 
4\mu ^{2}A_{s}^{4}\Delta T,n\neq 1%
\end{array}%
\right. ,  \label{e6}
\end{equation}%
where $\Delta T$ is the sequence period.

\begin{figure}[tbph]
\centering\includegraphics[width=5.047in]{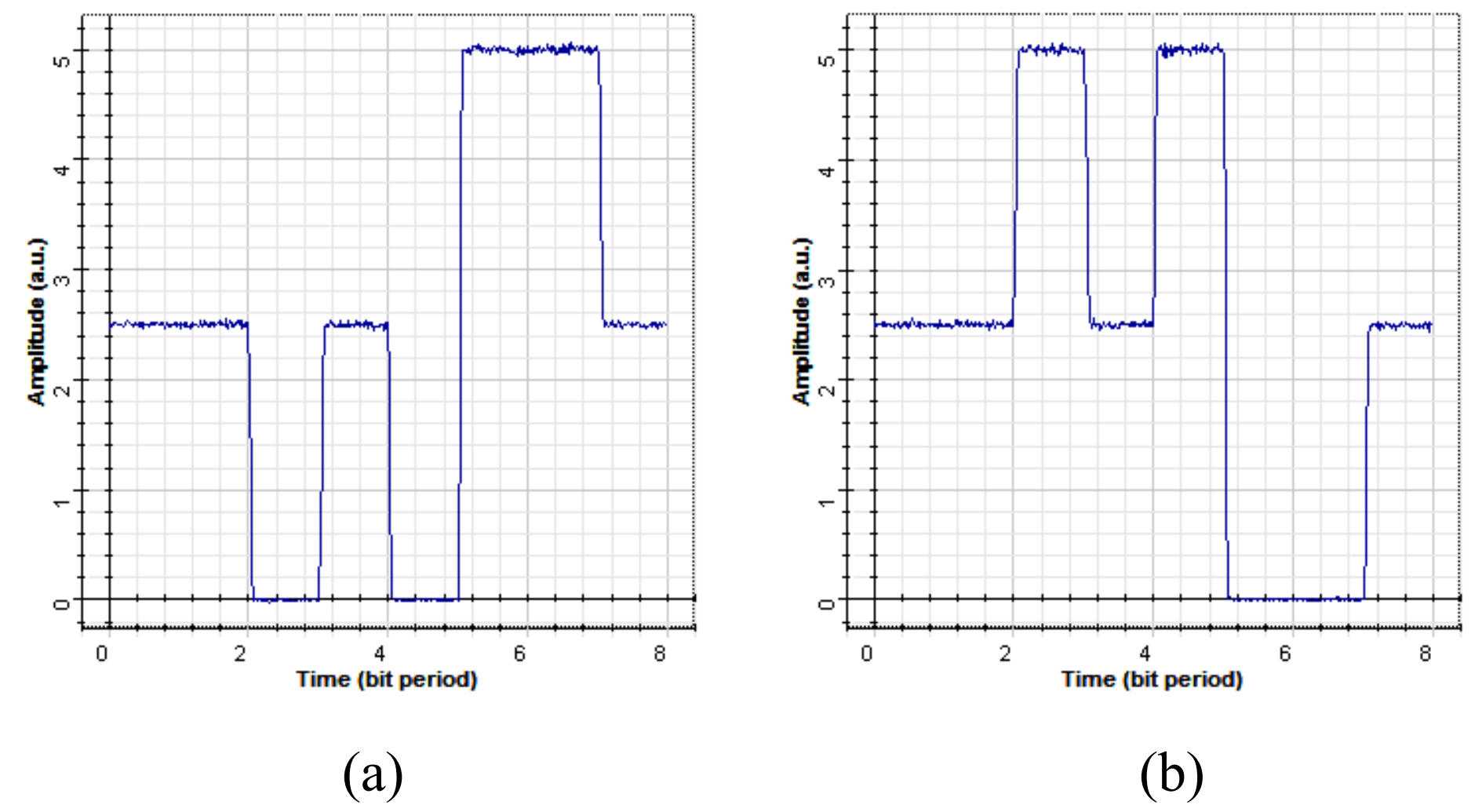}
\caption{The electric signals of $D_{1}$(a) and $D_{2}$(b) when the sequence
of LO is $\protect\lambda ^{\left( 5\right) }$.}
\label{2}
\end{figure}

\begin{figure}[tbph]
\centering\includegraphics[width=5.047in]{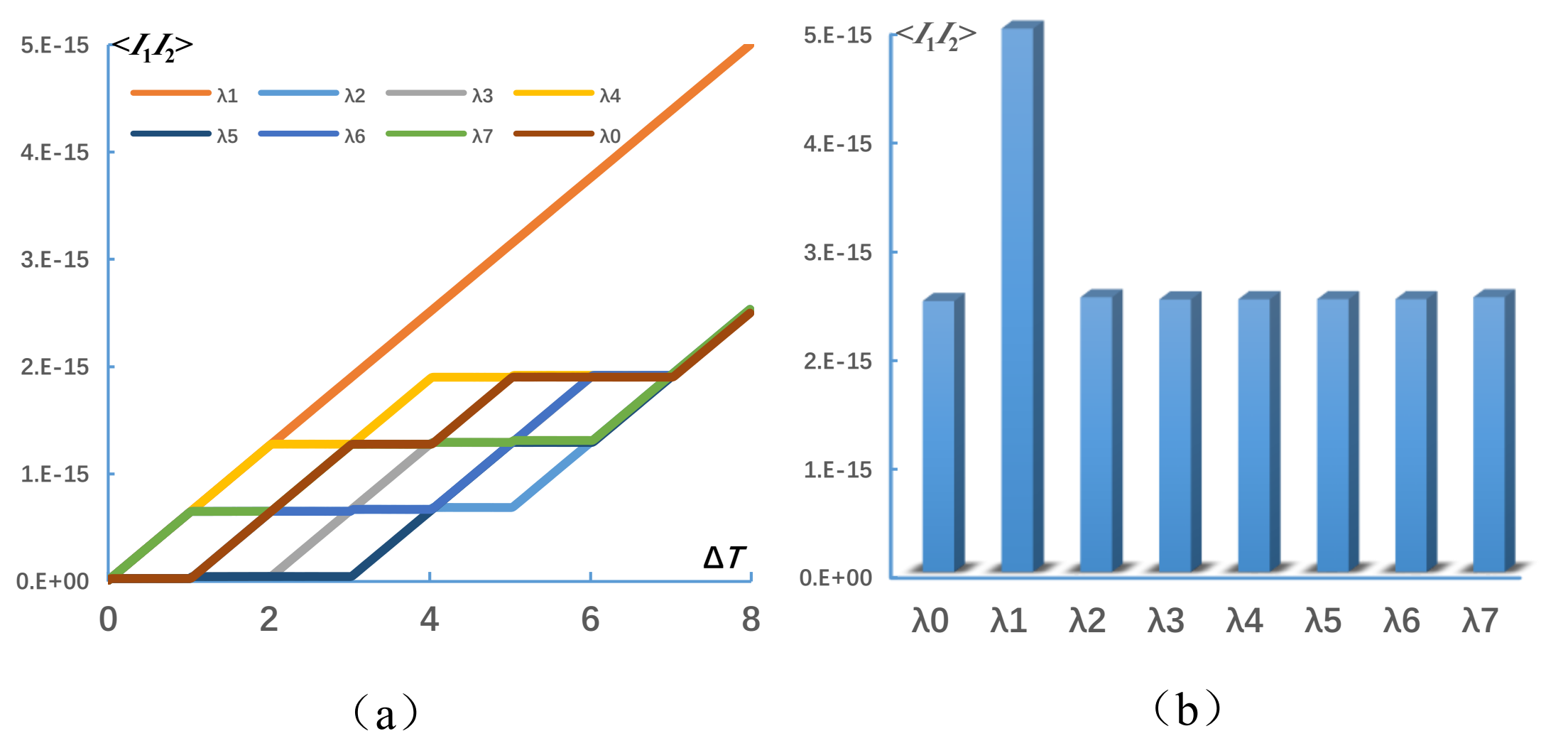}
\caption{The correlation analysis result between signal light and local
light modulated with different pseudorandom sequence, where (a) is the
intergral of correlation function with different sequence period $\Delta T$,
and (b) is the final result of intergral of different LO with $\protect%
\lambda 0\sim \protect\lambda 7$ represent sequences $\protect\lambda %
^{\left( 0\right) }\sim \protect\lambda ^{\left( 7\right) }$.}
\label{3}
\end{figure}

To verify the above scheme, we utilize OPTISYSTEM to simulate it on the
computer. Fig. \ref{2} shows the electric signals of two PDs within a
sequence period. Fig. \ref{3} shows the result after the correlation of the
optical fields modulate with different sequences, from which we can find out
when the modulation sequences of local light and signal light is same, the
value of correlation function is one time larger than that in other cases.
Hence, the orthogonality of pseudorandom sequences can be used to
distinguish the optical fields modulated by different phase sequences. By
using the polarization beam splitter, we can easily realize the detection of
the classical fields with two polarization modes.

\section{\textbf{Analogy to} the three-particle quantum states \label{sec3}}

\begin{figure}[tbph]
\centering\includegraphics[width=3.2491in]{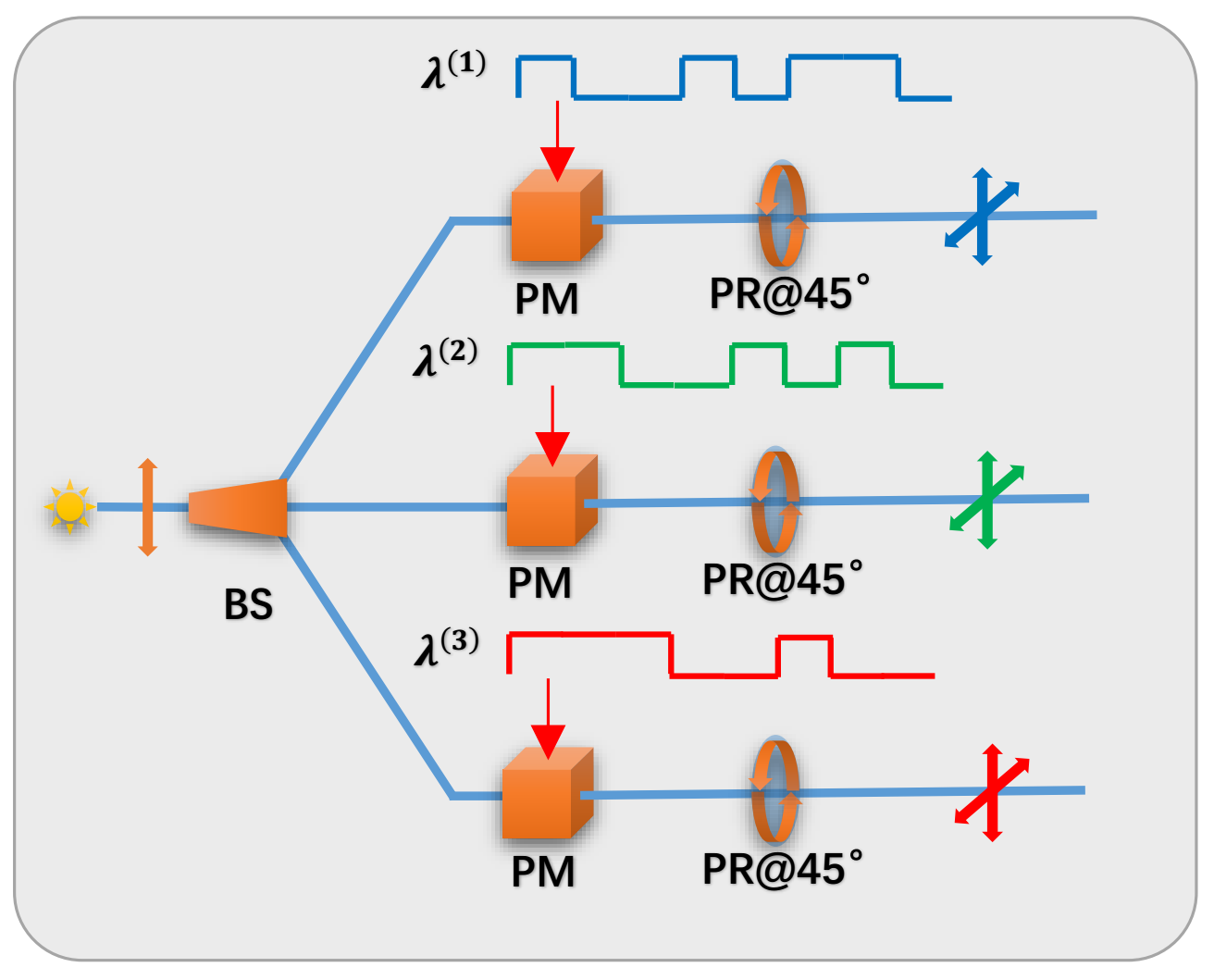}
\caption{The sheme to realize the simulation of quantum product state, where
PR@$45^{\circ }$: $45^{\circ }$ polarization rotators.}
\label{4}
\end{figure}

Ref. \cite{Fu2} demostarates the classical field simulation of
three-particle quantum states, which are a product state, GHZ state and W
state. To construct product state, we arbitrarily choose three sequences
from the pseudorandom sequences set to modulate the classical fields, and
obtain: 
\begin{eqnarray}
E_{1}\left( t\right) &=&\left( A_{\uparrow }+A_{\rightarrow }\right)
e^{-i\left( \omega t+\lambda _{k}^{\left( 1\right) }\right) },  \label{e7} \\
E_{2}\left( t\right) &=&\left( A_{\uparrow }+A_{\rightarrow }\right)
e^{-i\left( \omega t+\lambda _{k}^{\left( 2\right) }\right) },  \nonumber \\
E_{3}\left( t\right) &=&\left( A_{\uparrow }+A_{\rightarrow }\right)
e^{-i\left( \omega t+\lambda _{k}^{\left( 3\right) }\right) },  \nonumber
\end{eqnarray}%
where $A_{\uparrow }$ and $A_{\rightarrow }$ dnote the amplitudes of two
orthogonal polarization modes $\left\vert 0\right\rangle $ and $\left\vert
1\right\rangle $, respectively. We adopt the scheme in Fig. \ref{4} to
realize this state.

According to Ref. \cite{Fu2}, the classical field simulations of GHZ state
can be written as following:%
\begin{eqnarray}
E_{1}\left( t\right) &=&A_{\uparrow }e^{-i\left( \omega t+\lambda
_{k}^{\left( 1\right) }\right) }+A_{\rightarrow }e^{-i\left( \omega
t+\lambda _{k}^{\left( 2\right) }\right) },  \label{e8} \\
E_{2}\left( t\right) &=&A_{\uparrow }e^{-i\left( \omega t+\lambda
_{k}^{\left( 2\right) }\right) }+A_{\rightarrow }e^{-i\left( \omega
t+\lambda _{k}^{\left( 3\right) }\right) },  \nonumber \\
E_{3}\left( t\right) &=&A_{\uparrow }e^{-i\left( \omega t+\lambda
_{k}^{\left( 3\right) }\right) }+A_{\rightarrow }e^{-i\left( \omega
t+\lambda _{k}^{\left( 1\right) }\right) },  \nonumber
\end{eqnarray}%
which can be realized by mode exchange of the produce state in equation (\ref%
{e7}) using polarization beam splitters (PBSs), as shown in Fig. \ref{6}.
Then we can express the classical field simulations of W state as following: 
\begin{eqnarray}
E_{1}\left( t\right) &=&A_{\uparrow }e^{-i\left( \omega t+\lambda
_{k}^{\left( 1\right) }\right) }+A_{\rightarrow }e^{-i\left( \omega
t+\lambda _{k}^{\left( 2\right) }\right) }+A_{\rightarrow }e^{-i\left(
\omega t+\lambda _{k}^{\left( 3\right) }\right) },  \label{e9} \\
E_{2}\left( t\right) &=&A_{\uparrow }e^{-i\left( \omega t+\lambda
_{k}^{\left( 1\right) }\right) }+A_{\rightarrow }e^{-i\left( \omega
t+\lambda _{k}^{\left( 2\right) }\right) }+A_{\rightarrow }e^{-i\left(
\omega t+\lambda _{k}^{\left( 3\right) }\right) },  \nonumber \\
E_{3}\left( t\right) &=&A_{\uparrow }e^{-i\left( \omega t+\lambda
_{k}^{\left( 1\right) }\right) }+A_{\rightarrow }e^{-i\left( \omega
t+\lambda _{k}^{\left( 2\right) }\right) }+A_{\rightarrow }e^{-i\left(
\omega t+\lambda _{k}^{\left( 3\right) }\right) },  \nonumber
\end{eqnarray}%
which can be realized by mode combination and split of the initial state
using the beam coupler and splitter, as shown in Fig. \ref{7}.

\begin{figure}[tbph]
\centering\includegraphics[width=5.047in]{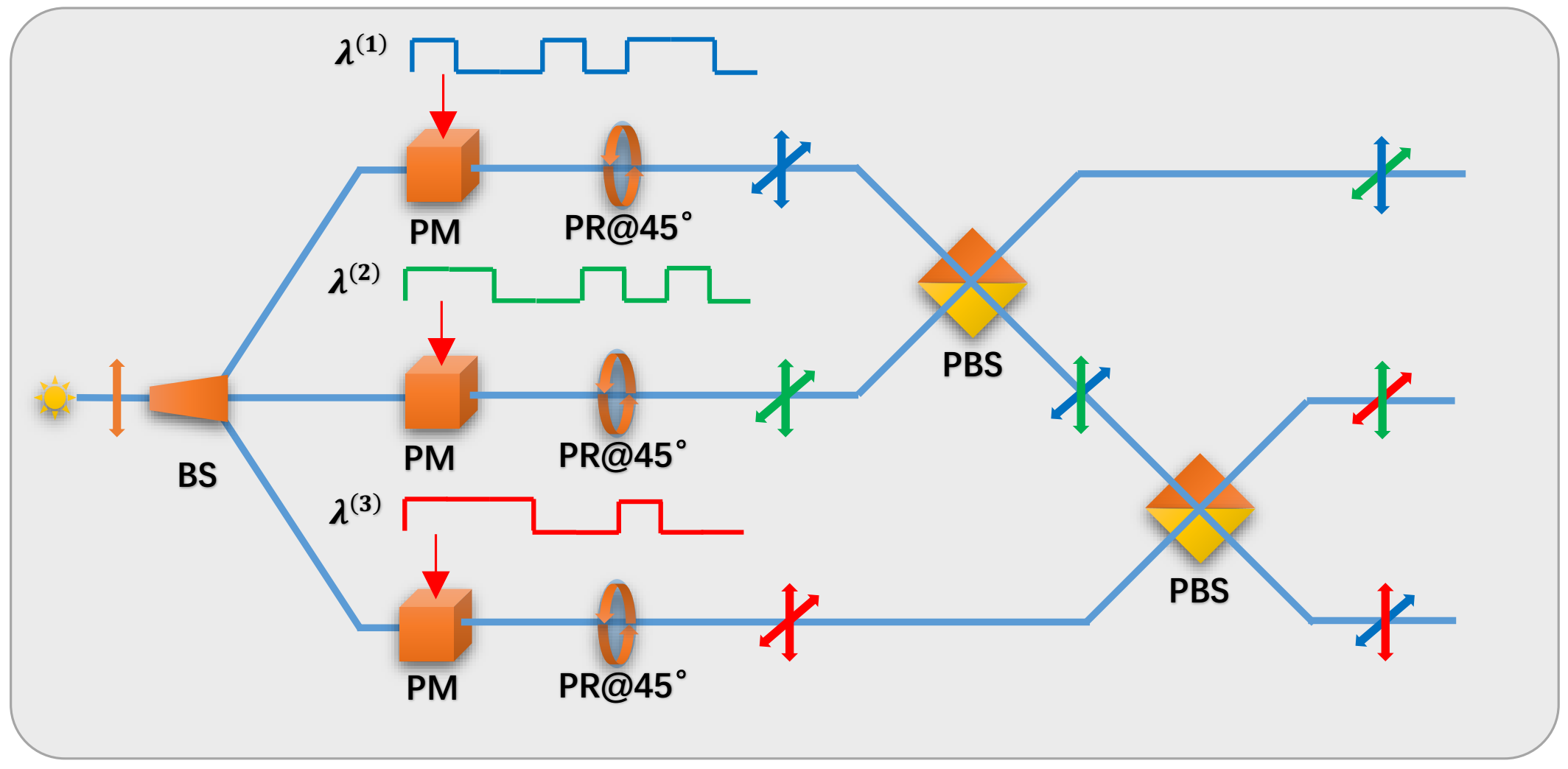}
\caption{The scheme to realize the simulation of quantum GHZ state, where
PBS: polarization beam splitter, PR@$45^{\circ }$: $45^{\circ }$
polarization rotators.}
\label{5}
\end{figure}

\begin{figure}[tbph]
\centering\includegraphics[width=5.047in]{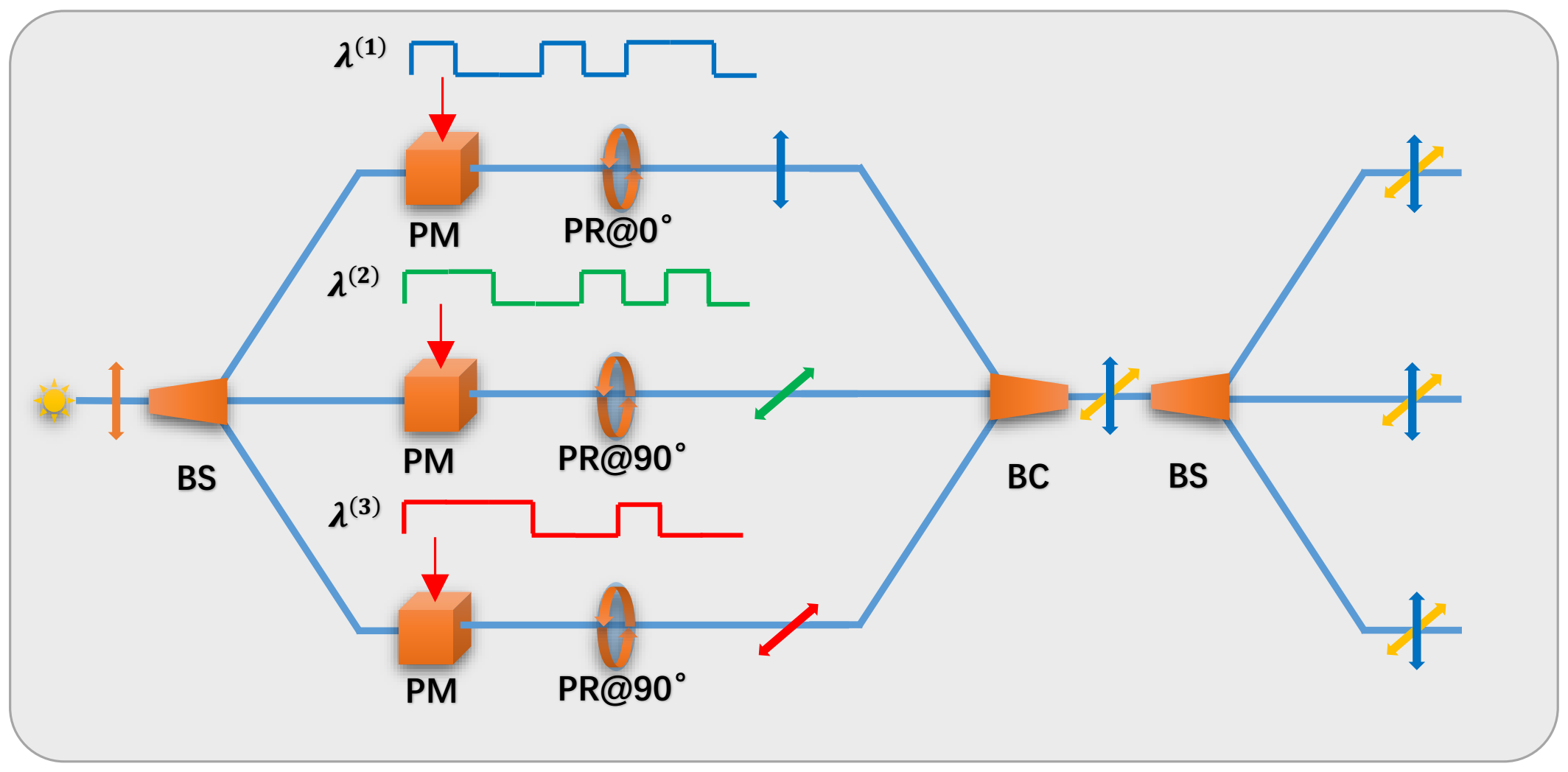}
\caption{The sheme to realize the simulation of quantum W state, where BC:
beam coupler, BS: beam splitter, PR@$0^{\circ }$: $0^{\circ }$ polarization
rotators, PR@$90^{\circ }$: $90^{\circ }$ polarization rotators.}
\label{6}
\end{figure}

Here, we make use of the method mentioned in Section \ref{sec2} to
investigate the relation between the coherent detection of pseudorandom
sequence and the simulation of quantum states. Because of each field of GHZ
state and W state has two orthogonal polarization modes, the detection
scheme need to split two modes using a PBS, as shown in Fig. \ref{7}. By
using OPTISYSTEM, we can easily constructe a simulation model to realize the
schemes as shown in Fig. \ref{5} and Fig. \ref{7} for GHZ state. Fig. \ref{8}
shows the electric signals of PDs after the interference between the first
field and LO, where the LOs are modulated by three pseudorandom sequences $%
\lambda ^{\left( 1\right) }$, $\lambda ^{\left( 2\right) }$, $\lambda
^{\left( 3\right) }$ respectively, and (a) and (b) represent two orthogonal
polarization modes respectively. Then, we can obtain the correlation
function of three fields, as shown in Fig. \ref{9}. After substracting the
correlation function with constant and normalization, we can express the
measurement result as the $M$ matrix mentioned in Ref. \cite{Fu2} as
following:

\begin{equation}
M=\left( 
\begin{array}{ccc}
\left( 1,0\right) & \left( 0,1\right) & 0 \\ 
0 & \left( 1,0\right) & \left( 0,1\right) \\ 
\left( 0,1\right) & 0 & \left( 1,0\right)%
\end{array}%
\right) ,  \label{e10}
\end{equation}%
where the rows denote classical fields $E_{1}\left( t\right) ,E_{2}\left(
t\right) $ and $E_{3}\left( t\right) $, the columns denote the pseudorandom
sequences $\lambda ^{\left( 1\right) }$, $\lambda ^{\left( 2\right) }$, $%
\lambda ^{\left( 3\right) }$, and the matrix elements denote the states of
mode ($(1,0)$ denotes $A_{\uparrow }$ exists, $(0,1)$ denotes $%
A_{\rightarrow }$ exists, $(1,1)$ denotes all modes exist, $0$ denote none
mode exists).

\begin{figure}[tbph]
\centering\includegraphics[width=5.047in]{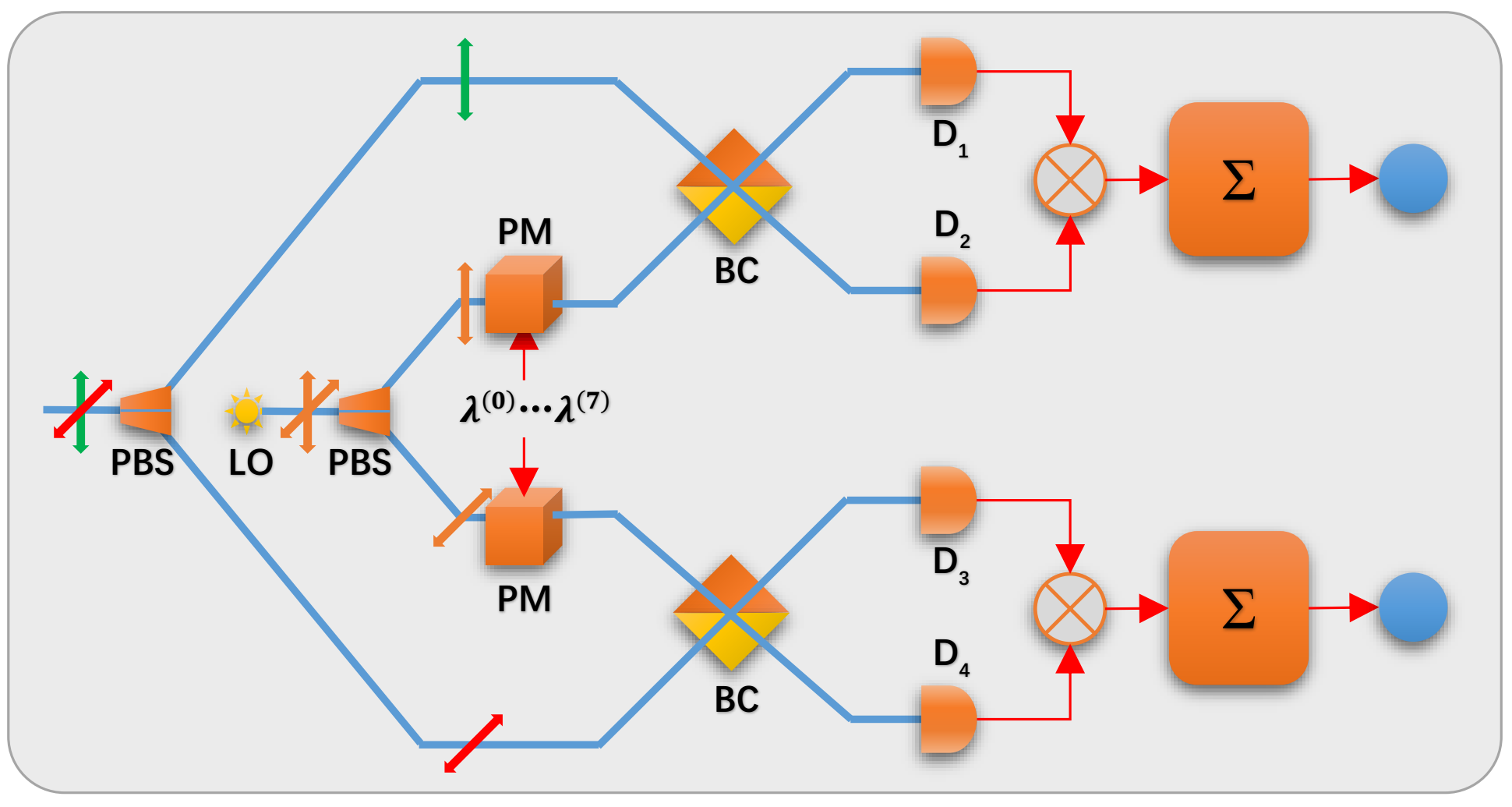}
\caption{The coherent detection scheme of GHZ state and W state, where PBS:
polarization beam splitter and BC: beam coupler.}
\label{7}
\end{figure}

\begin{figure}[tbph]
\centering\includegraphics[width=5.047in]{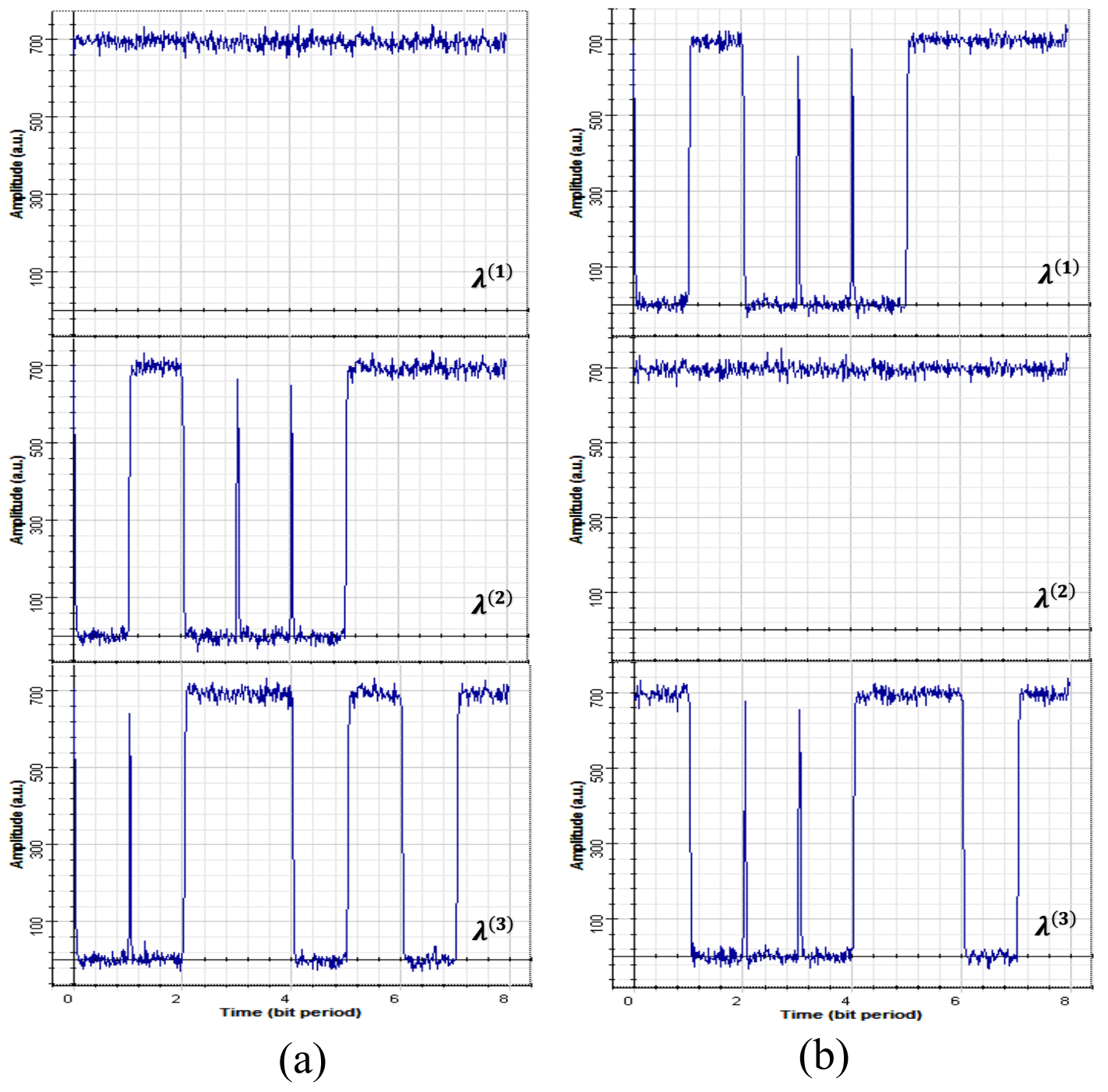}
\caption{The electric signal of the coherent detection of the first field of
GHZ state, where (a) and (b) represent two orthogonal modes $A_{\uparrow }$
and $A_{\rightarrow }$ respectively.}
\label{8}
\end{figure}

\begin{figure}[tbph]
\centering\includegraphics[width=5.047in]{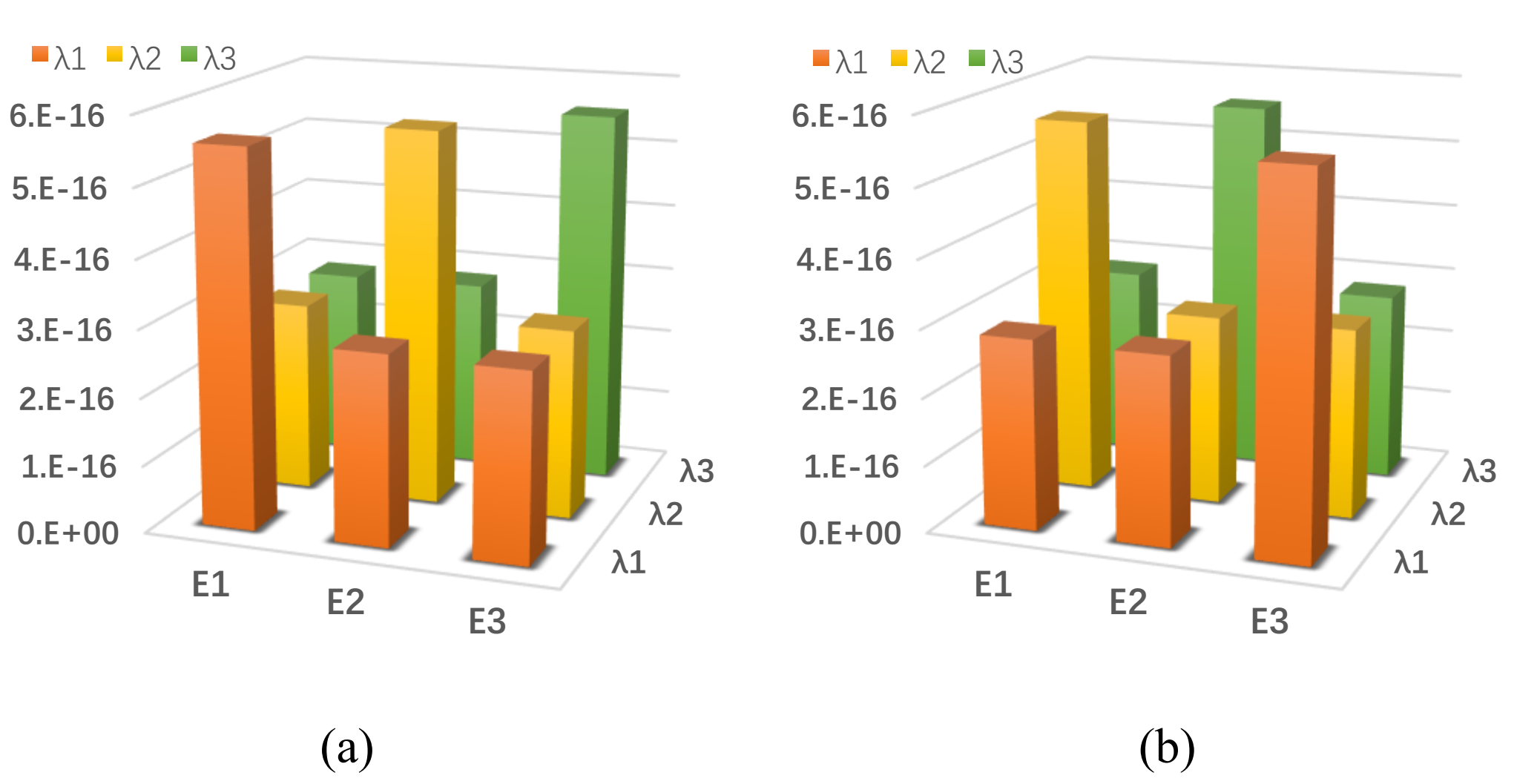}
\caption{The correlation measurement result of GHZ state, where (a) and (b)
represent two orthogonal modes $A_{\uparrow }$ and $A_{\rightarrow }$
respectively, $E1$, $E2$, $E3$ three fields and $\protect\lambda 1$, $%
\protect\lambda 2$, $\protect\lambda 3$ the consequences $\protect\lambda %
^{\left( 1\right) }$, $\protect\lambda ^{\left( 2\right) }$, $\protect%
\lambda ^{\left( 3\right) }$ modulating on LO.}
\label{9}
\end{figure}

\begin{figure}[tbph]
\centering\includegraphics[width=5.047in]{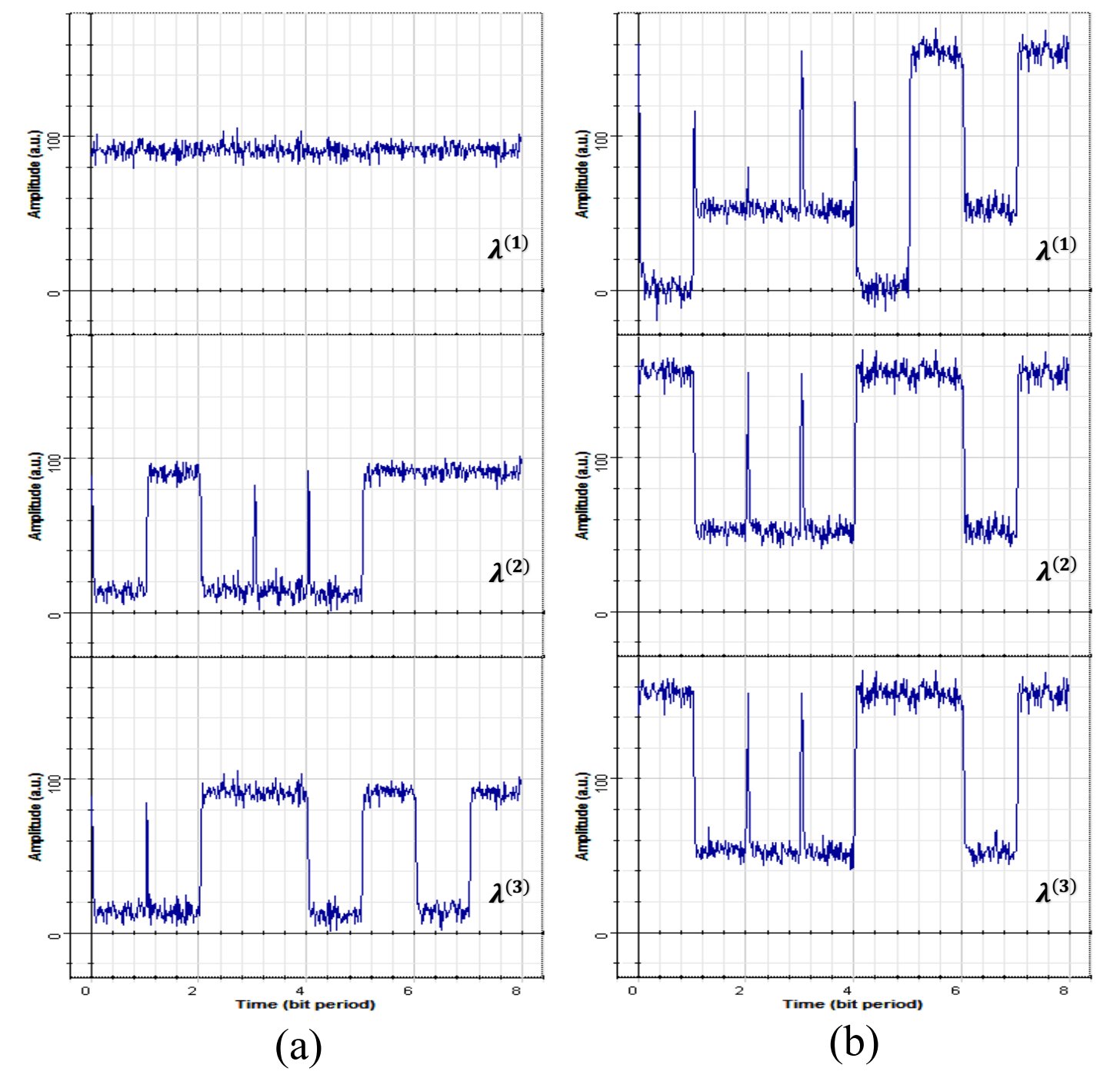}
\caption{The electric signal of the coherent detection of the first field of
W state, where (a) and (b) represent two orthogonal modes $A_{\uparrow }$
and $A_{\rightarrow }$ respectively.}
\label{10}
\end{figure}

\begin{figure}[tbph]
\centering\includegraphics[width=5.047in]{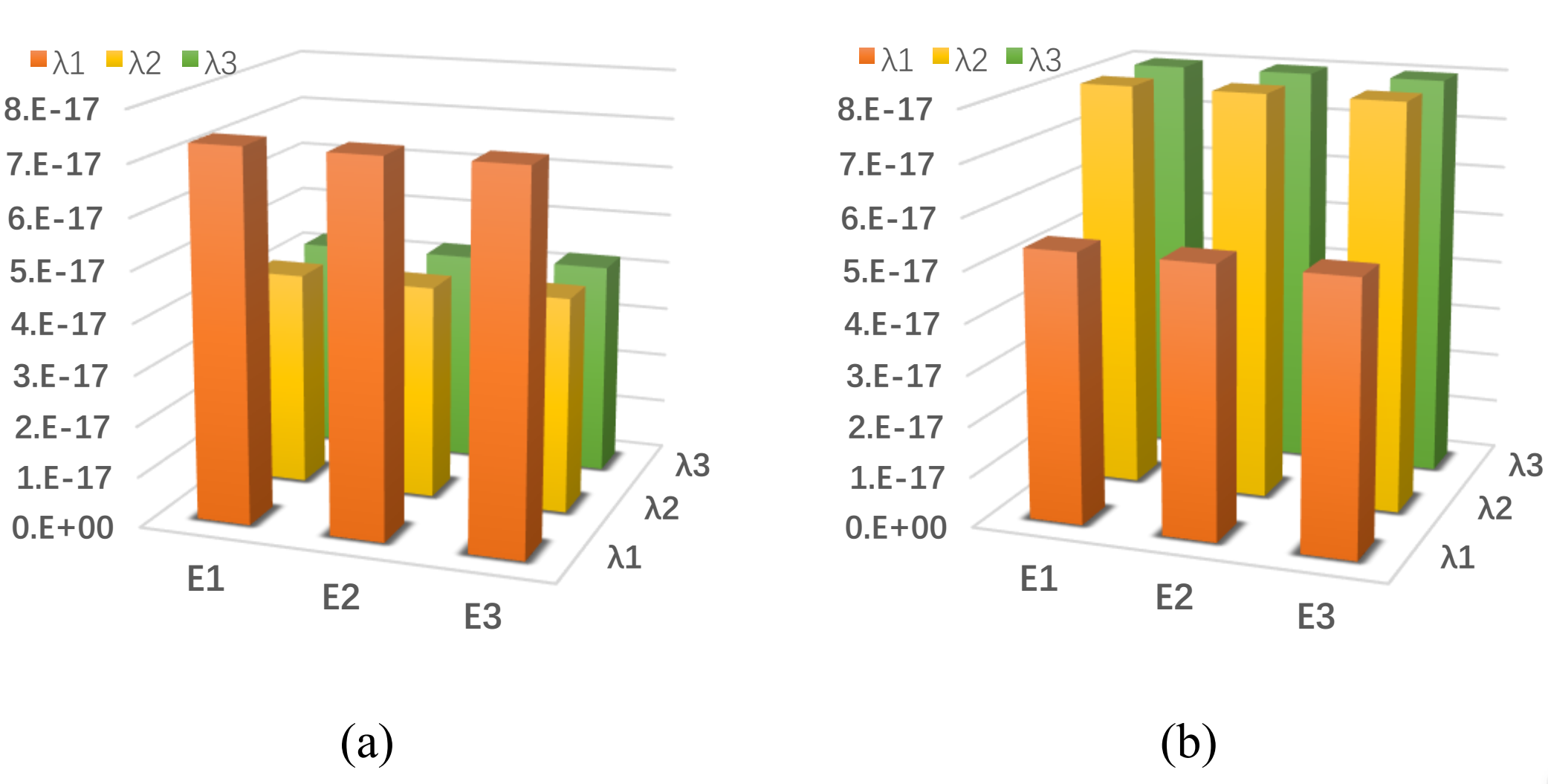}
\caption{The correlation measurement result of W state, where (a) and (b)
represent two orthogonal modes $A_{\uparrow }$ and $A_{\rightarrow }$
respectively, $E1$, $E2$, $E3$ three fields and $\protect\lambda 1$, $%
\protect\lambda 2$, $\protect\lambda 3$ the consequences $\protect\lambda %
^{\left( 1\right) }$, $\protect\lambda ^{\left( 2\right) }$, $\protect%
\lambda ^{\left( 3\right) }$ modulating on LO.}
\label{11}
\end{figure}

Meanwhile, we can simulate W state. Fig. \ref{10} show the electric signals
of PDs after the interference between the first field of W state and LO.
Furthermore, we can obtain the correlation function of three fields, as
shown in Fig. \ref{11}. After substracting the correlation function with
constant and normalization, we can express the $M$ matrix of W state as
following:%
\begin{equation}
M=\left( 
\begin{array}{ccc}
\left( 1,0\right) & \left( 0,1\right) & \left( 0,1\right) \\ 
\left( 1,0\right) & \left( 0,1\right) & \left( 0,1\right) \\ 
\left( 1,0\right) & \left( 0,1\right) & \left( 0,1\right)%
\end{array}%
\right) .  \label{e11}
\end{equation}

\section{Analogy to the result state for factorizing $15=3\times 5$ \label%
{sec4}}

\begin{figure}[tbph]
\centering\includegraphics[width=5.3169in]{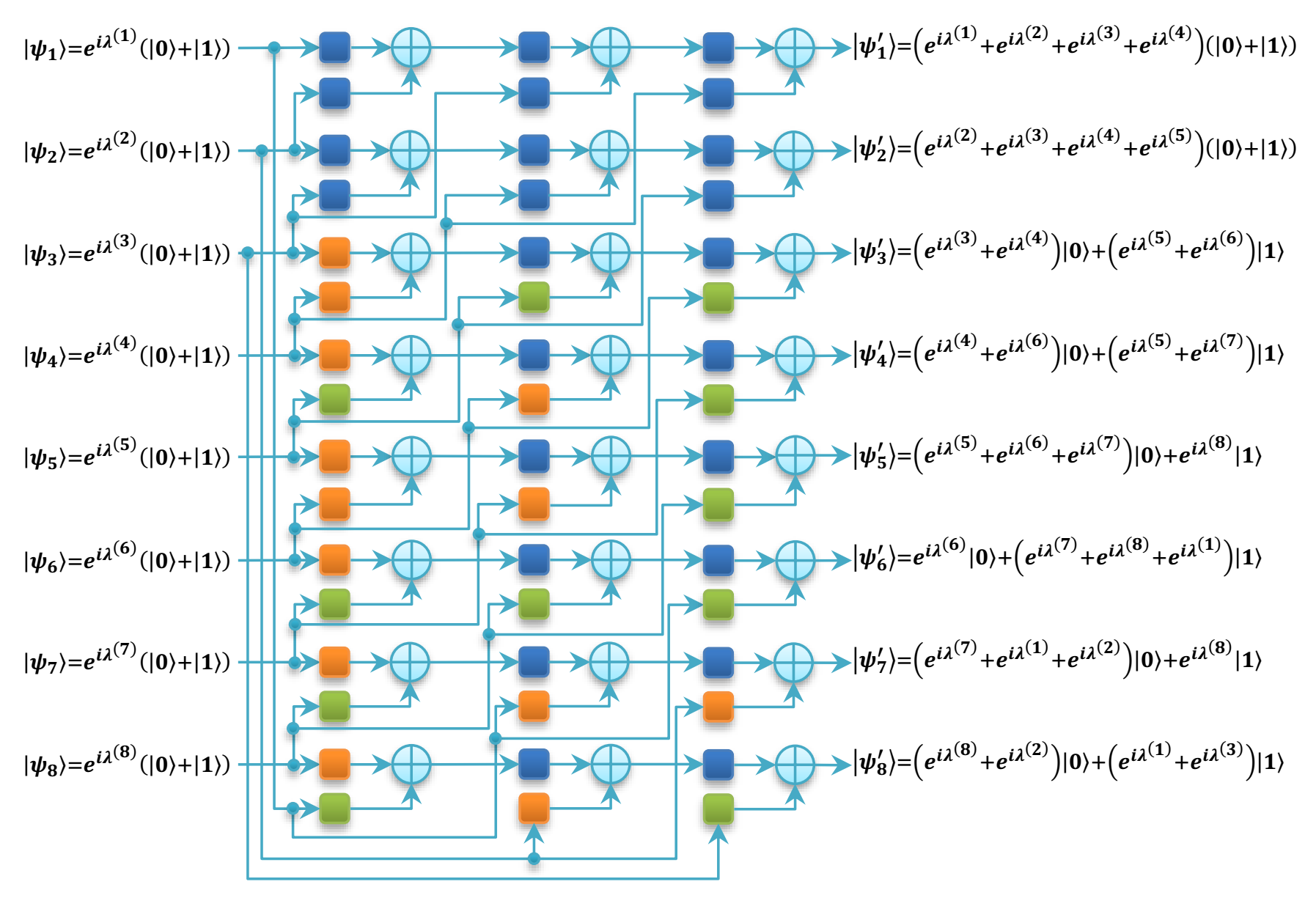}
\caption{Mode transformation gate matrix, where the blue block denotes all
modes pass, the orange block $A_{\uparrow }$ mode ($\left\vert
0\right\rangle $) passes, and the green block $A_{\rightarrow }$ mode passes
($\left\vert 1\right\rangle $).}
\label{12}
\end{figure}

Shor's algorithm is a crucial algorithm displaying the exponential speed-up
of quantum computation. Shor's algorithm to factorize $15=3\times 5$ has
been verified by NMR experiment \cite{Vandersypen}. The key to Shor's
algorithm to factorize positive integer $N$ is to calculate the modular
exponential function $f(x)=a^{x}modN$, where $a$ is positive coprime
integer, and obtain the result state that is an eight-particle entangled
state, then to apply the fourier transformation on the state to obtain the
period of it. According to Ref. \cite{Fu2}, we can obtain optical analogy to
the entangled state as the result of the modular exponential function. The
classical fields are modulated by $8$ pseudorandom sequences. After a series
of the polarization operations as shown in Fig. \ref{12}, the required state
can be obtained,

\begin{eqnarray}
\left\vert \psi _{1}^{\prime }\right\rangle &=&\left( e^{i\lambda ^{\left(
1\right) }}+e^{i\lambda ^{\left( 2\right) }}+e^{i\lambda ^{\left( 3\right)
}}+e^{i\lambda ^{\left( 4\right) }}\right) \left( \left\vert 0\right\rangle
+\left\vert 1\right\rangle \right) ,  \label{eq12} \\
\left\vert \psi _{2}^{\prime }\right\rangle &=&\left( e^{i\lambda ^{\left(
2\right) }}+e^{i\lambda ^{\left( 3\right) }}+e^{i\lambda ^{\left( 4\right)
}}+e^{i\lambda ^{\left( 5\right) }}\right) \left( \left\vert 0\right\rangle
+\left\vert 1\right\rangle \right) ,  \nonumber \\
\left\vert \psi _{3}^{\prime }\right\rangle &=&\left( e^{i\lambda ^{\left(
3\right) }}+e^{i\lambda ^{\left( 4\right) }}\right) \left\vert
0\right\rangle +\left( e^{i\lambda ^{\left( 5\right) }}+e^{i\lambda ^{\left(
6\right) }}\right) \left\vert 1\right\rangle ,  \nonumber \\
\left\vert \psi _{4}^{\prime }\right\rangle &=&\left( e^{i\lambda ^{\left(
4\right) }}+e^{i\lambda ^{\left( 6\right) }}\right) \left\vert
0\right\rangle +\left( e^{i\lambda ^{\left( 5\right) }}+e^{i\lambda ^{\left(
7\right) }}\right) \left\vert 1\right\rangle ,  \nonumber \\
\left\vert \psi _{5}^{\prime }\right\rangle &=&\left( e^{i\lambda ^{\left(
5\right) }}+e^{i\lambda ^{\left( 6\right) }}+e^{i\lambda ^{\left( 7\right)
}}\right) \left\vert 0\right\rangle +e^{i\lambda ^{\left( 8\right)
}}\left\vert 1\right\rangle ,  \nonumber \\
\left\vert \psi _{6}^{\prime }\right\rangle &=&e^{i\lambda ^{\left( 6\right)
}}\left\vert 0\right\rangle +\left( e^{i\lambda ^{\left( 7\right)
}}+e^{i\lambda ^{\left( 8\right) }}+e^{i\lambda ^{\left( 1\right) }}\right)
\left\vert 1\right\rangle ,  \nonumber \\
\left\vert \psi _{7}^{\prime }\right\rangle &=&\left( e^{i\lambda ^{\left(
7\right) }}+e^{i\lambda ^{\left( 1\right) }}+e^{i\lambda ^{\left( 2\right)
}}\right) \left\vert 0\right\rangle +e^{i\lambda ^{\left( 8\right)
}}\left\vert 1\right\rangle ,  \nonumber \\
\left\vert \psi _{8}^{\prime }\right\rangle &=&\left( e^{i\lambda ^{\left(
8\right) }}+e^{i\lambda ^{\left( 2\right) }}\right) \left\vert
0\right\rangle +\left( e^{i\lambda ^{\left( 1\right) }}+e^{i\lambda ^{\left(
3\right) }}\right) \left\vert 1\right\rangle .  \nonumber
\end{eqnarray}

\begin{figure}[tbph]
\centering\includegraphics[width=5.3177in]{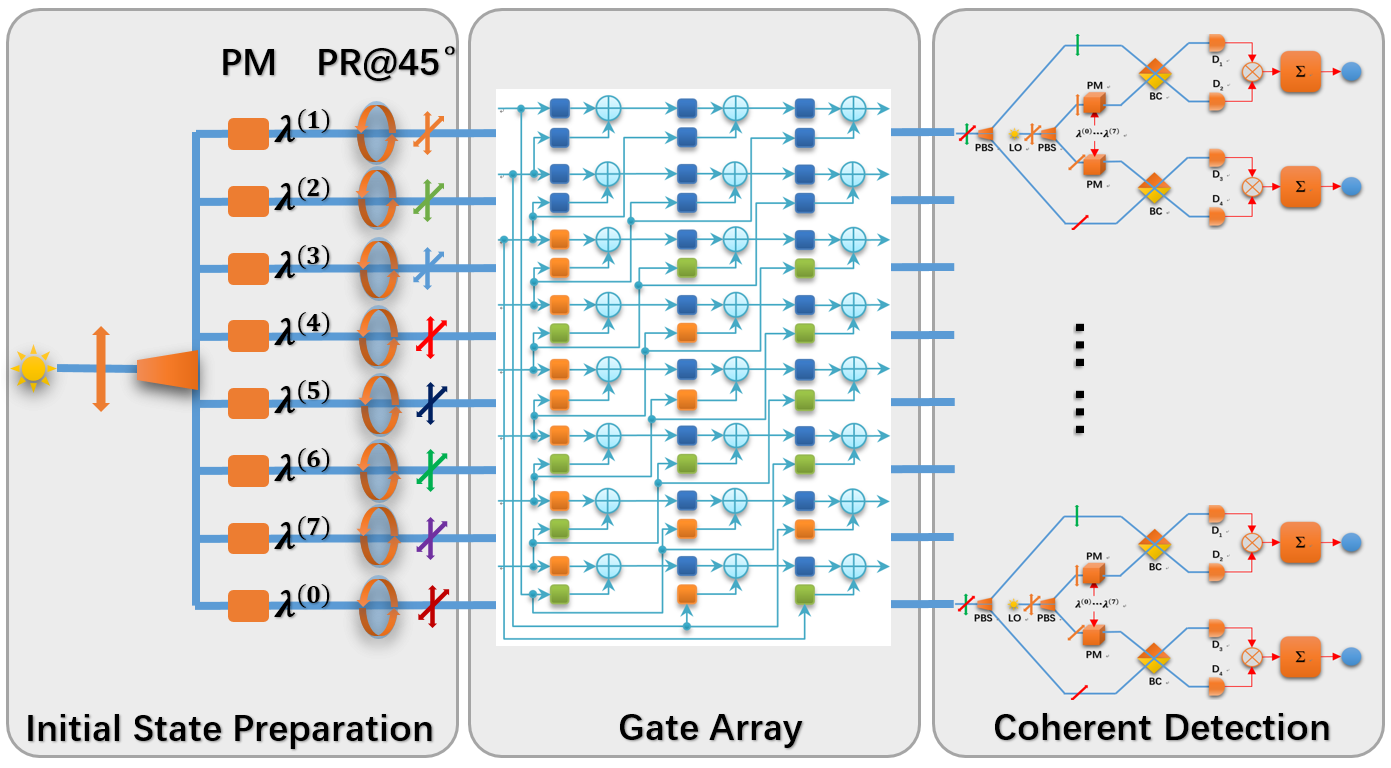}
\caption{The computer simulation scheme of algorithm factorizing $15=3\times
5$.}
\label{13}
\end{figure}

\begin{figure}[tbph]
\centering\includegraphics[width=5.3177in]{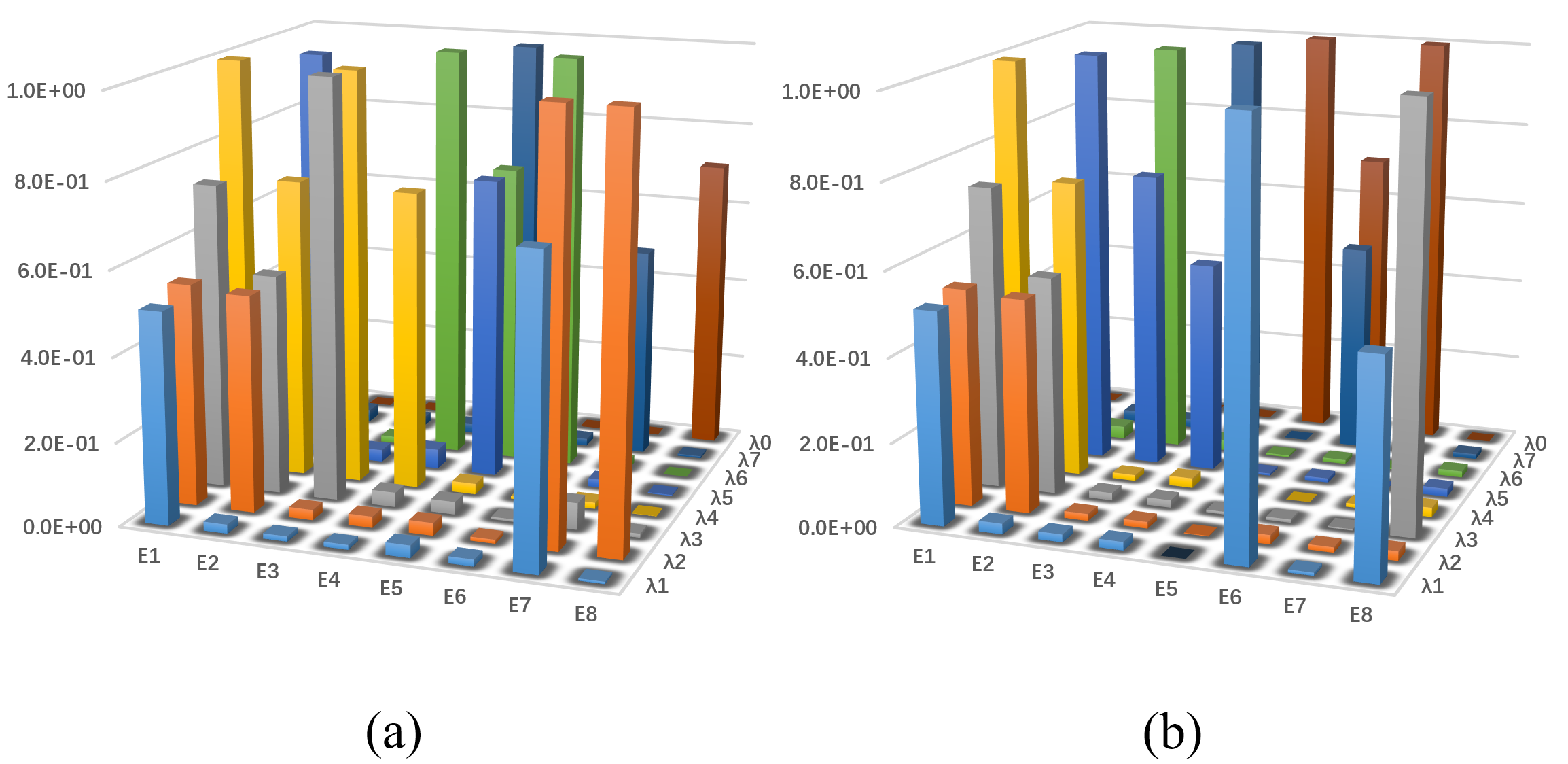}
\caption{The correlation measurement result of the superposition state of
the factorizing algorithm: (a) for mode $A_{\uparrow }$, and (b) for mode $%
A_{\rightarrow }$, where $E1\sim E8$ represents $E_{1}\symbol{126}E_{8}$
classical fields, and $\protect\lambda 0\sim \protect\lambda 7$ the
pseudorandom sequences modulated on LO $\protect\lambda ^{\left( 0\right)
}\sim \protect\lambda ^{\left( 7\right) }.$}
\label{14}
\end{figure}

By using OPTISYSTEM, we constructe a simulation model to realize the schemes
as shown Fig. \ref{12} to obtain the result of the modular exponential
function. Then the final result fields are coherently detected and the $M$
matrix can be obtained. The schematic diagram of computer simulation is
shown in Fig. \ref{13}. Firstly, the initial state can be prepared by the
method as shown in Fig. \ref{4}, that is, modulate $8$ classical fields $%
E_{1}\symbol{126}E_{8}$ with $8$ pseudorandom sequences in Eq. (\ref{e1})
respectively, and then rotate the polarization of each field by $45^{\circ }$
and evolve into mode superposition states. After numerically simulating a
complex gate array, we obtain the output fields and electric signals of PDs
after the interference between each fields and LO. Finally, the correlation
results are obtained, substracted by the constant part and normalized, as
shown in Fig. \ref{14}. After the threshold discrimination and binarization
of results, we can express the measurement results as the $M$ matrix, which
is: 
\begin{equation}
M=\left( 
\begin{array}{cccccccc}
\left( 1,1\right) & \left( 1,1\right) & \left( 1,1\right) & \left( 1,1\right)
& 0 & 0 & 0 & 0 \\ 
0 & \left( 1,1\right) & \left( 1,1\right) & \left( 1,1\right) & \left(
1,1\right) & 0 & 0 & 0 \\ 
0 & 0 & \left( 1,0\right) & \left( 1,0\right) & \left( 0,1\right) & \left(
0,1\right) & 0 & 0 \\ 
0 & 0 & 0 & \left( 1,0\right) & \left( 0,1\right) & \left( 1,0\right) & 
\left( 0,1\right) & 0 \\ 
0 & 0 & 0 & 0 & \left( 1,0\right) & \left( 1,0\right) & \left( 1,0\right) & 
\left( 0,1\right) \\ 
\left( 0,1\right) & 0 & 0 & 0 & 0 & \left( 1,0\right) & \left( 0,1\right) & 
\left( 0,1\right) \\ 
\left( 1,0\right) & \left( 1,0\right) & 0 & 0 & 0 & 0 & \left( 1,0\right) & 
\left( 0,1\right) \\ 
\left( 0,1\right) & \left( 1,0\right) & \left( 0,1\right) & 0 & 0 & 0 & 0 & 
\left( 1,0\right)%
\end{array}%
\right)  \label{eq13}
\end{equation}%
Using the sequence permutation scheme mentioned in Ref. \cite{Fu2}, we can
obtain the simulated states: 
\begin{eqnarray}
\left\vert \Psi ^{\prime }\right\rangle &=&\left( \left\vert 0\right\rangle
+\left\vert 4\right\rangle +\left\vert 8\right\rangle +\left\vert
12\right\rangle \right) \left\vert 1\right\rangle  \label{eq14} \\
&&+\left( \left\vert 1\right\rangle +\left\vert 5\right\rangle +\left\vert
9\right\rangle +\left\vert 13\right\rangle \right) \left\vert 7\right\rangle
\nonumber \\
&&+\left( \left\vert 2\right\rangle +\left\vert 6\right\rangle +\left\vert
10\right\rangle +\left\vert 14\right\rangle \right) \left\vert 4\right\rangle
\nonumber \\
&&+\left( \left\vert 3\right\rangle +\left\vert 7\right\rangle +\left\vert
11\right\rangle +\left\vert 15\right\rangle \right) \left\vert
13\right\rangle .  \nonumber
\end{eqnarray}

There are four kinds of superposition classified from last four qubits
containing the values of $f\left( x\right) $ ($\left\vert 1\right\rangle
,\left\vert 7\right\rangle ,\left\vert 4\right\rangle $ and $\left\vert
13\right\rangle $) in output states, which means the period of $f\left(
x\right) =7^{x}mod15$ is $r=4$. It is worth noting that, different
from quantum computing, we obtain the expected period of without operating
quantum Fourier transformation \cite{Fu4}.

\section{Conclusions \label{sec5}}

In this paper, we utilize numerical simulations to prove the feasibility of
coherent detection and correlation analysis to distinguish mixing classical
fields with pseudorandom phase sequences. Due to much bigger Hilbert space
spanned by the classical fields than quantum states \cite{Fu1}, we might
realize optical analogies to any quantum states. A rigorous proof will be
discussed in the future paper.

\begin{description}
\item \newpage

\item[Fig. 1] The scheme of the coherent detection of pseudorandom phase
sequence, where SO: the signal light, LO: the local light, BS: beam
splitter, PM: phase modulator, BC: beam coupler, $D_{1}$ and $D_{2}$:
photodetectors, $\otimes $: multiplier and $\Sigma $: integrator (integrate
over entire sequence period).

\item[Fig. 2] The electric signals of $D_{1}$(a) and $D_{2}$(b) when the
sequence of LO is $\lambda ^{\left( 5\right) }$.

\item[Fig. 3] The correlation analysis result between signal light and local
light modulated with different pseudorandom sequence, where (a) is the
intergral of correlation function with different sequence period $\Delta T$,
and (b) is the final result of intergral of different LO with $\lambda 0\sim
\lambda 7$ represent sequences $\lambda ^{\left( 0\right) }\sim \lambda
^{\left( 7\right) }$.

\item[Fig. 4] The sheme to realize the simulation of quantum product state,
where PR@$45^{\circ }$: $45^{\circ }$ polarization rotators.

\item[Fig. 5] The scheme to realize the simulation of quantum GHZ state,
where PBS: polarization beam splitter, PR@$45^{\circ }$: $45^{\circ }$
polarization rotators.

\item[Fig. 6] The sheme to realize the simulation of quantum W state, where
BC: beam coupler, BS: beam splitter, PR@$0^{\circ }$: $0^{\circ }$
polarization rotators, PR@$90^{\circ }$: $90^{\circ }$ polarization rotators.

\item[Fig. 7] The coherent detection scheme of GHZ state and W state, where
PBS: polarization beam splitter and BC: beam coupler.

\item[Fig. 8] The electric signal of the coherent detection of the first
field of GHZ state, where (a) and (b) represent two orthogonal modes $%
A_{\uparrow }$ and $A_{\rightarrow }$ respectively.

\item[Fig. 9] The correlation measurement result of GHZ state, where (a) and
(b) represent two orthogonal modes $A_{\uparrow }$ and $A_{\rightarrow }$
respectively, $E1$, $E2$, $E3$ three fields and $\lambda 1$, $\lambda 2$, $%
\lambda 3$ the consequences $\lambda ^{\left( 1\right) }$, $\lambda ^{\left(
2\right) }$, $\lambda ^{\left( 3\right) }$ modulating on LO.

\item[Fig. 10] The electric signal of the coherent detection of the first
field of W state, where (a) and (b) represent two orthogonal modes $%
A_{\uparrow }$ and $A_{\rightarrow }$ respectively.

\item[Fig. 11] The correlation measurement result of W state, where (a) and
(b) represent two orthogonal modes $A_{\uparrow }$ and $A_{\rightarrow }$
respectively, $E1$, $E2$, $E3$ three fields and $\lambda 1$, $\lambda 2$, $%
\lambda 3$ the consequences $\lambda ^{\left( 1\right) }$, $\lambda ^{\left(
2\right) }$, $\lambda ^{\left( 3\right) }$ modulating on LO.

\item[Fig. 12] Mode transformation gate matrix, where the blue block denotes
all modes pass, the orange block $A_{\uparrow }$ mode ($\left\vert
0\right\rangle $) passes, and the green block $A_{\rightarrow }$ mode passes
($\left\vert 1\right\rangle $).

\item[Fig. 13] The computer simulation scheme of algorithm factorizing $%
15=3\times 5$.

\item[Fig. 14] The correlation measurement result of the superposition state
of factorizing algorithm: (a) for mode $A_{\uparrow }$, and (b) for mode $%
A_{\rightarrow }$, where $E1\sim E8$ represents $E_{1}\sim E_{8}$ classical
fields, and $\lambda 0\sim \lambda 7$ the pseudorandom sequences modulated
on LO $\lambda ^{\left( 0\right) }\sim \lambda ^{\left( 7\right) }.$
\end{description}

\end{document}